\documentclass{article}
\usepackage{spconf,amsmath,graphicx,url,subfig, cite,booktabs}
\usepackage{amsfonts,amssymb}
\usepackage{xcolor}
\usepackage{makecell}
\usepackage[hidelinks]{hyperref}
\usepackage{svg}
\usepackage{multirow}
\usepackage{booktabs} 

\title{TWO-STEP BAND-SPLIT NEURAL NETWORK APPROACH FOR FULL-BAND  \\ RESIDUAL ECHO suppression}

\name{
\begin{tabular}{c}
\it Zihan Zhang$^1$, Shimin Zhang$^1$, Mingshuai Liu$^1$, Yanhong Leng$^2$, Zhe Han$^2$, Li Chen$^2$, Lei Xie$^{1,*}$\thanks{$^*$: Corresponding author.}
\end{tabular}
\vspace{-0.2cm} 
}
\address{
  $^1$Audio, Speech and Language Processing Group (ASLP@NPU), School of Computer Science, \\ Northwestern Polytechnical University, Xi'an, China\\
  $^2$ByteDance, China \\
  }

\begin{document}
\ninept
\maketitle
\begin{abstract}
\vspace{-4pt}


This paper describes a Two-step Band-split Neural Network (TBNN) approach for full-band acoustic echo cancellation. Specifically, after linear filtering, we split the full-band signal into wide-band (16KHz) and high-band (16-48KHz) for residual echo removal with lower modeling difficulty. The wide-band signal is processed by an updated gated convolutional recurrent network (GCRN) with U$^2$ encoder while the high-band signal is processed by a high-band post-filter net with lower complexity. Our approach submitted to ICASSP 2023 AEC Challenge has achieved an overall mean opinion score (MOS) of 4.344 and a word accuracy (WAcc) ratio of 0.795, leading to the 2$^{nd}$ (tied) in the ranking of the non-personalized track.

\end{abstract}

\vspace{-2pt}

\vspace{-12pt}
\section{Introduction}
\label{sec:intro}
\vspace{-8pt}

The 4th acoustic echo cancellation (AEC) challenge~\cite{aec2023icassp} is a flagship event of the ICASSP 2023 signal processing grand challenge, with the aim to benchmark AEC techniques in real-time full-band (48KHz) speech communication. In this challenge, our team
has submitted a \textit{hybrid} approach that combines a linear filter with a neural post-filter to the non-personalized AEC track (i.e., without using target speaker embedding). 
Since noise-like components usually dominate in higher bands of speech and structured harmonics are mainly found in wide-band signals, we propose a two-step band-split neural network (TBNN) approach to particularly handle full-band residual echo removal on wide-band (16KHz) and high-band (16-48KHz) in a two-step process, as an extension of our previous work~\cite{zhang2022multi} to better model full-band signals with low complexity. Specifically, the wide-band poster filter (WBPF) is based on the gated convolutional recurrent network (GCRN)~\cite{zhang2022multi} but with upgraded U$^2$ encoder~\cite{qin2020u2} for better latent feature extraction. We also redesign the data simulation method and the loss function to accommodate full-band signals. 
According to the results, our system has ranked 2nd place (tied rank) in the non-personalized track with an overall mean opinion score (MOS) of 4.344 and a word accuracy (WAcc) ratio of 0.795.

\vspace{-8pt}
\section{Proposed Method}
\label{sec:format}
\vspace{-6pt}
\subsection{Problem formulation}
\vspace{-6pt}
For a typical full-duplex communication system consisting of a microphone and a loudspeaker, the signal model can be expressed as
\begin{equation}
\footnotesize
\label{eq1} d(n)=s(n)+r(n)+v(n)+z(n)
\end{equation}
\begin{equation}
\footnotesize
\label{eq2} z(n)=h(n) \ast \mathcal{F}\{x(n)\}
\end{equation}
where the microphone signal $d(n)$ is composed of the near-end speech signal $s(n)$ which may include early reflections, late reverberation $r(n)$, additive noise $v(n)$, and echo signal $z(n)$, where $n$ is the sampling index. 
The echo signal $z(n)$ is obtained by convolving the echo path $h(n)$ with the nonlinear distorted reference signal $x(n)$, shown in Eq.~(\ref{eq2}), where $\mathcal{F}$ refers to the nonlinear distortions, $*$ refers to convolution. An AEC system aims to cancel $z(n)$ from $d(n)$ given $x(n)$. For complex real-life scenarios, noise $v(n)$ and reverberation $r(n)$ also need to be suppressed.

\begin{figure*}[!ht]
    \centering
    \includegraphics[scale=0.57]{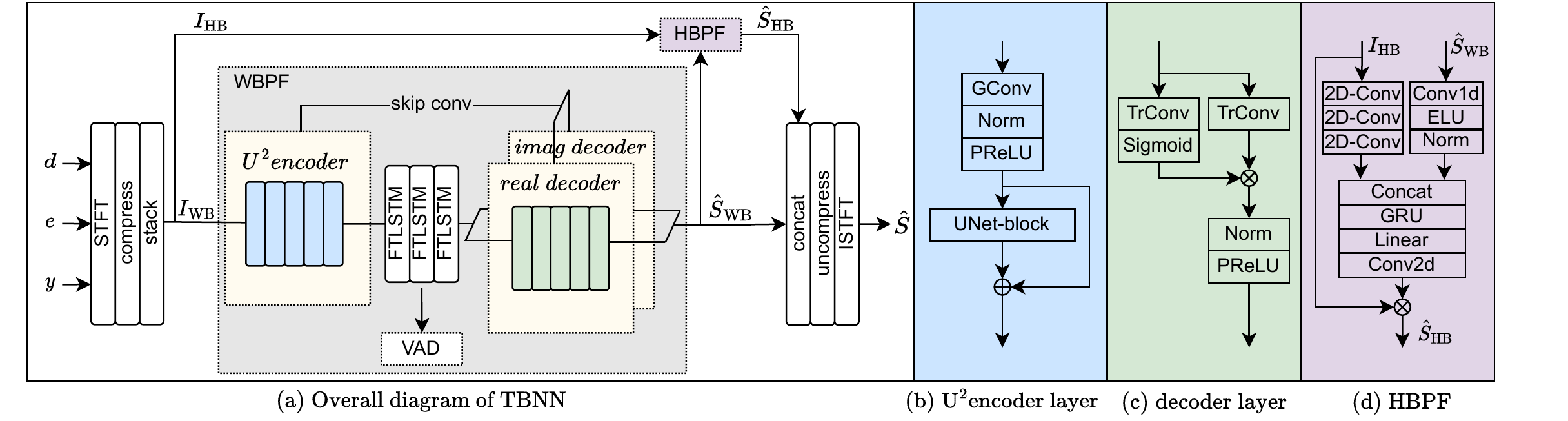}\vspace{-0.3cm}
    \caption{The proposed TBNN post-filter.}\vspace{-0.6cm}
    \label{fig:TGCRN}
\end{figure*}
We adopt a time delay estimation (TDE) module based on the sub-band cross-correlation to align the reference signal. Our linear filter uses a sub-band adaptive filtering method based on the NLMS algorithm to estimate the linear echo signal $y$ and the error signal $e$.
\vspace{-14pt}
\subsection{TBNN post-filter}
\vspace{-6pt}
The TBNN post-filter takes $d$, $e$, and $y$ after a short-time Fourier transform (STFT) as input. The complex-valued input spectra are compressed by a factor of 0.5, stacked by real and imaginary parts, and divided into wide-band signal $I_{\text{WB}}$ and high-band signal $I_{\text{HB}}$.
Considering the complexity of direct full-band modeling and the high-band has less structural information, we use a two-step band-split approach to model full-band signals.
The wide-band post-filter (WBPF) first uses spectral mapping to estimate the wide-band output $\hat{S}_{\text{WB}}$. Then the high-band post-filter (HBPF) uses $I_{\text{HB}}$ and priori information $\hat{S}_{\text{WB}}$ to predict the complex-valued mask $M_{\text{HB}}$, which is multiplied by $I_{\text{HB}}$ to obtain the high-band output $\hat{S}_{\text{HB}}$.

The backbone of the WBPF module is based on our previous gated convolutional recurrent network (GCRN)~\cite{zhang2022multi}. But differently, the encoder consists of 5 $\text U^{2}\text{-Encoder}$ layers as shown in Fig.~\ref{fig:TGCRN}(b). $\text U^{2}\text{-Net}$~\cite{qin2020u2} is a two-level nested U-structure designed to capture more contextual information from different scales (through Residual U-blocks) without significantly increasing the computational cost. Specifically, each $\text U^{2}\text{-Encoder}$ layer is composed of a gated convolution (GConv) layer, a batch-norm (BN) layer, a PReLU and an UNet-block. The number of input and output channels remains the same. The FTLSTM layer is served as a bottleneck for temporal modeling, as defined in~\cite{zhang21ia_interspeech}. The decoder consists of 5 decoder blocks as shown in Fig.~\ref{fig:TGCRN}(c), where TrConv refers to the Transpose-Conv2d layer, and $\otimes$ refers to multiply.
The skip connection (implemented with 1×1 convolution) is applied between the encoder and decoder. 
We also add a voice activity detection (VAD) module to avoid over-suppression of near-end speech~\cite{zhang2022multi}.

The HBPF module uses three 2D-Conv modules to extract high-band features. Each 2D-Conv module consists of a 2D convolution (Conv2d) layer, an ELU layer, a BN layer and a dropout layer with a 0.25 dropout rate~\cite{zhang2022two}. 
The wide-band output serves as prior information and it is concatenated with the high-band features after feature dimension alignment. Then we use the GRU layer and Conv2d layer to predict the real/imaginary mask, which is applied to the high-band input to obtain the high-band output.

\vspace{-14pt}
\subsection{Loss function}
\vspace{-4pt}
The loss function is composed of an echo power weighted magnitude loss and a power-law compressed phase aware (PLCPA) loss, our previous work has proved the effectiveness of echo-weighted loss~\cite{zhang2022multi}.
The difference is that we calculate the loss separately for wide-band and high-band.
A VAD loss and a masked mean square error (MSE) loss are added as auxiliaries to avoid near-end speech over-suppression. Thus for the wide-band part, the loss is
{
\setlength\abovedisplayskip{2.0pt}
\setlength\belowdisplayskip{2.0pt}
\begin{equation}
\mathcal{L}_\text{WB}=\mathcal{L}_\text{echo-weighted} + \mathcal{L}_\text{plcpa} + 0.5\cdot\mathcal{L}_\text{mask} + 0.1\cdot\mathcal{L}_\text{vad}.
\end{equation}
}
For the high-band prediction, the loss is
{
\setlength\abovedisplayskip{2.0pt}
\setlength\belowdisplayskip{2.0pt}
\begin{equation}
\mathcal{L}_\text{HB}=\mathcal{L}_\text{echo-weighted} + \mathcal{L}_\text{plcpa} + 0.5 \cdot\mathcal{L}_\text{mask}.
\end{equation}
}
The final loss function is
{
\vspace{-2pt}
\setlength\abovedisplayskip{2.0pt}
\setlength\belowdisplayskip{2.0pt}
\begin{equation} \label{eq10}
\mathcal{L}_\text{final} = \alpha  \mathcal{L}_\text{HB} + \mathcal{L}_\text{WB}.
\end{equation}
}
Due to the dynamic range difference between high-band and wide-band, we empirically set $\alpha=10$.

\vspace{-15pt}
\section{Experiments}
\vspace{-9pt}
\subsection{Dataset}
\vspace{-6pt}
The near-end signals are taken from the 4th deep noise suppression (DNS) challenge, and the noises are taken from Freesound and AudioSet. The echo dataset is composed of real recordings and synthetic data. Real recordings are taken from the AEC-Challenge far-end single-talk dataset. We remove the recordings that contain near-end speech, in order to avoid model over-suppression on near-end speech. Synthetic data is generated by convolving room impulse responses (RIRs) with the DNS data. A total of 50,000 RIRs are generated using the HYB method\cite{9306409}, with random room size 5×3×3 to 8×5×4 and rt60 0.2-1.2s. Totally 30\% of the near-end speech is convoluted with RIRs to simulate reverbed signals. The total training data has 2000 hours of real-recorded echo and 1000 hours of simulated echo, while SNR randomly ranges from 0dB to 25 dB and the SER randomly ranges from -15dB to 15 dB. 
The validation and test sets are generated in the same way as above, with 50 hours and 1500 clips, respectively. 

\vspace{-12pt}
\subsection{Experimental setup}
\vspace{-6pt}
Window length and frame shift are 20 ms and 10 ms, respectively and 960-point STFT is applied. For each Conv2d, we use a convolution kernel of (2, 3) and a stride of (1, 2). The FTLSTM configuration is the same as~\cite{zhang21ia_interspeech}. We experiment with two model sizes -- the number of Conv2d channels in the encoder/decoder is 80 for the smaller one and 128 for the large one (with -L in the model name). We take our previous GCRN~\cite{zhang2022multi} for comparison, which is a wide-band model without HBPF and thus our previous subband method~\cite{zhang2022multi} is adopted to process full-band signal. U$^{2}$-based encoder is also tested for GCRN, named as GCRN-U$^{2}$. For the proposed TBNN model, the 2D-conv block in the HBPF module has an output channel of 128, the 1D convolution layer (Conv1d) is a point-wise convolution with an output channel of 48, and the GRU layer has a hidden state of 256. We particularly compare the performance of \textit{masking} and \textit{mapping} in HBPF.


\begin{table}[]
\scriptsize
\centering
 \caption{Echo suppression performance. ST-FE: far-end single-talk, ST-NE: near-end single-talk, DT: double-talk}
 \vspace{-8pt}
 \label{tab:corrcoef}
  \resizebox{\linewidth}{!}{
\begin{tabular}{@{}lccccc@{}}
\toprule
    & Para. (M) & \makecell[c]{ST-FE\\(ERLE)} & \makecell[c]{DT\\(WB-PESQ)} & \makecell[c]{ST-NE\\(WB-PESQ)} & Data \\ \midrule
Input &  -  & 0   & 1.76 & 2.36 &    \\ \midrule
GCRN & 2.45 & 52.10 & 2.41 &  2.68 & \multirow{6}{*}{400h}   \\ 
GCRN-L & 7.31 &  57.29 & 2.48 & 2.78 &   \\ 
$\text{GCRN-U}^{2}$ & 3.69 & 57.21 &  2.50 &  2.81 &   \\ \cmidrule(r){0-4} 
$\text{TBNN}\text{-mapping}$ & 5.54 & 58.58 & 2.57 & 2.87 &   \\
$\text{TBNN}\text{-masking}$ & 5.45 & 57.10 & 2.60 & 2.92 &   \\ 
$\text{TBNN}\text{-L-masking}$ & 9.56 & 59.99 &  2.62 & 2.93 &   \\ \midrule
$\text{TBNN}\text{-L-masking}$ & 9.56 & \textbf{63.12} & \textbf{2.73} & \textbf{3.06} &  3000h \\
\bottomrule
\end{tabular}
}
\vspace{-8pt}
\end{table}

\begin{table}[]
\small
\centering
\vspace{-3pt}
 \caption{Evaluation results on blind test set}
 \vspace{-8pt}
 \label{tab:results}
  \resizebox{\linewidth}{!}{
\begin{tabular}{@{}lccccccc@{}}
\toprule
    & \makecell[c]{ST-FE\\MOS}  & \makecell[c]{DT-ECHO\\MOS} & \makecell[c]{DT-Other\\MOS} & \makecell[c]{ST-NE\\SIG MOS} & \makecell[c]{ST-NE\\BAK MOS} & \makecell[c]{Overall\\MOS} & WAcc \\ \midrule
baseline &  4.535 & 4.283   & 3.479 & 3.883 & 3.887 & 4.013 & 0.649  \\ 
$\text{TBNN}\text{-L-masking}$ & \textbf{4.704} & \textbf{4.725} & \textbf{4.169} & \textbf{3.918} &  \textbf{4.202} & \textbf{4.344} & \textbf{0.795} \\
\bottomrule
\end{tabular}
}
\vspace{-17pt}
\end{table}

\vspace{-10pt}
\subsection{Results and conclusions}
\vspace{-6pt}

We use a 400-hour subset for quick experimentation. As shown in Table~\ref{tab:corrcoef}, We have the following conclusions. First, the use of $\text{U}^{2}\text{-Encoder}$ can achieve better performance with much fewer parameters compared to simply increasing the channel number ($\text{GCRN-U}^{2}$ vs. GCRN-L). Second, the proposed TBNN is better than GCRN which mainly focuses on the wide-band signal. Third, masking in HBPF leads to better near-end speech quality as compared with mapping. Finally, by enlarging the model size, TBNN-L-masking obtains the best performance. Thus the model trained with the entire 3000 hours of data is used to process the blind test set as our submission and results are shown in Table~\ref{tab:results}. Our submitted model achieved 2nd place (tied) in the challenge ranking. Its real-time factor (RTF) is 0.35, tested on Intel(R) Xeon(R) E5-2678 v3 @2.50GHz using a single thread, comprising 0.28 for the TBNN post-filter (exported by ONNX) and 0.07 for the time delay estimation and adaptive filter.






\footnotesize
\vspace{-6pt}
\bibliographystyle{IEEEbib}
\let\oldbibliography\thebibliography 
\renewcommand{\thebibliography}[1]{ 
  \oldbibliography{#1}
  \setlength{\itemsep}{-1pt} 
}
\bibliography{strings,refs}

\end{document}